\documentclass[preprintnumbers, prd, onecolumn, floatfix,
preprintnumbers,showpacs, letterpaper, superscriptaddress,nofootinbib,11pt]{revtex4}
\usepackage{graphicx,epsfig}
\usepackage{amsmath}
\usepackage{ulem}
\usepackage{amsfonts}
\usepackage{amssymb}
\pdfoutput=1
\usepackage{amsmath, amssymb, graphicx}
\usepackage{epsf}
\usepackage{epstopdf}
\usepackage {amssymb}
\usepackage{array}
\usepackage[colorlinks]{hyperref}
\usepackage[usenames,dvipsnames]{color}
\definecolor{beamer@PRD}{RGB}{46,48,146}
\definecolor{oxfordblue}{rgb}{0.0, 0.13, 0.28}
\definecolor{burgundy}{rgb}{0.5, 0.0, 0.13}
\definecolor{darkolivegreen}{rgb}{0.33, 0.42, 0.18}
\definecolor{bluer}{rgb}{0.00,0.50,0.75}
\definecolor{richcarmine}{rgb}{0.84, 0.0, 0.25}
\definecolor{darkgreen}{rgb}{0.00,0.50,0.25}
\definecolor{dark}{rgb}{0.00,0.55,0.55}{}
\hypersetup{
	breaklinks=true,
	pdfstartview={FitH},    % fits the width of the page to the window
	colorlinks=true,       % false: boxed links; true: colored links
	linkcolor=green,          % color of internal links
	citecolor=magenta,        % color of links to bibliography
	filecolor=magenta,      % color of file links
	urlcolor=oxfordblue,           % color of external links
	anchorcolor=red,      % Color for anchor text
	linktocpage=true
}
\newcommand{\nc}{\newcommand}
\nc{\ba}{\begin{eqnarray}}
	\nc{\ea}{\end{eqnarray}}
\newcommand\be{\begin{equation}}
	\newcommand\ee{\end{equation}}
\newcommand\bea{\begin{equation}}
	\newcommand\eea{\end{equation}}
\def\be {\begin{equation}}
	\def\ee {\end{equation}}
\def\bea {\begin{eqnarray}}
	\def\eea {\end{eqnarray}}
\def\bc {\begin{center}}
	\def\ec {\end{center}}
\def\nn {\nonumber}

\def\({\left(}
\def\){\right)}
\def\[{\left[}
\def\]{\right]}

\DeclareGraphicsExtensions{.jpg,.pdf,.eps}

\begin{document}
	
	\begin{flushright}
		%{\footnotesize IPM}
	\end{flushright}

	\title {\Large \bf Strongly Magnetized Hot QCD Matter and
		Stochastic Gravitational Wave Background}
	
	\author{\textbf{Mohsen Khodadi}}\email{m.khodadi@ipm.ir}
	\affiliation{School of Astronomy, Institute for Research in Fundamental Sciences (IPM),\\P.O. Box 19395-5531, Tehran, Iran}
	
	\author{\textbf{Ujjal Kumar Dey}}\email{ujjal@iiserbpr.ac.in}
	\affiliation{Department of Physical Sciences, Indian Institute of Science Education and Research Berhampur,\\Transit Campus, Government ITI, Berhampur 760010, Odisha, India}
	
	\author{\textbf{Gaetano Lambiase}}\email{lambiase@sa.infn.it}
	\affiliation{Dipartimento di Fisica ``E.R Caianiello", Universit$\grave{a}$ degli Studi di Salerno,\\Via Giovanni Paolo II, 132-84084 Fisciano (SA), Italy}
	\affiliation{Istituto Nazionale di Fisica Nucleare - Gruppo Collegato di Salerno - Sezione di Napoli,\\ Via Giovanni Paolo II, 132 - 84084 Fisciano (SA), Italy}
	
	\date{\today}
	
	\begin{abstract}
		The first-order phase transitions in the early Universe are one of the well-known sources which release the stochastic background of gravitational waves. In this paper, we study the contribution of an external static and strong magnetic field on the stochastic background of gravitational waves (GWs) expected during QCD phase transition. In the light of the strongly magnetized hot QCD equation of state which deviated from the ideal gas up to the one-loop approximation, we estimate two phenomenologically important quantities: peak frequency redshifted to today ($f_{\rm peak}$) and GW strain amplitude ($h^2 \Omega_{gw}$). 
		The trace anomaly induced by the magnetized hot QCD matter around the phase transition  generates the stochastic background of GW with peak frequencies lower than the ideal gas-based signal (around nHz). Instead, the strain amplitudes corresponding to the peak frequencies are of the same order of magnitude of the expected signal from ideal gas. This may be promising in the sense that although the strong magnetic field could mask the expected stochastic background of GWs by upgrading the frequency sensitivity of detectors in the future, the magnetized GW is expected to be identified.
		Faced with the projected reach of detectors EPTA, IPTA, and SKA, we find that for the tail of the magnetized GW signals there remains a mild possibility of detection as it can reach the projected sensitivity of SKA.
	\end{abstract}
	
	\maketitle
	%\newpage
	
	%%%%%%%%%%%%%%%%%%%%%%%%%%%%%%%%%%%%%
	\section{Introduction}
	%%%%%%%%%%%%%%%%%%%%%%%%%%%%%%%%%%%%%
	Almost a century after the inception of Einstein's general theory of relativity the physical existence of the gravitational waves (GW) was experimentally confirmed by the LIGO-VIRGO Collaborations in 2016 \citep{Abbott:2016blz, Abbott:2016nmj}.
	The GWs recorded by these two detectors were named \textit{GW150914} and \textit{GW151226} which was produced from binary black hole mergers. After that,  the LIGO recorded several other events such as \textit{GW170104}, \textit{GW170814} originating from binary stellar-mass black hole mergers \citep{Abbott:2017vtc, Abbott:2017oio} and \textit{GW170817}, which was from a binary neutron star inspiral \citep{TheLIGOScientific:2017qsa}. 
	The binary neutron star mergers may reveal richer physics since in addition to GWs it can emit electromagnetic signals which is not expected in the black hole merger\footnote{Recently for the first time in \citep{Graham:2020gwr} announced the existence of the electromagnetic signal to a binary black hole merger event \textit{S190521g}. However, the mentioned event has not yet been confirmed by the LIGO/Virgo collaboration.}. These early detections usher in a new chapter in astrophysics and cosmology research through a powerful tool that was not available to us before. We actually are at the beginning of the golden age of GW-based astronomy as a branch of observational astronomy which can play a complementary role to electromagnetic waves \citep{Sasaki:2018dmp}. In other words, GWs along with other exploration tools such as cosmic microwave background (CMB) and neutrinos are expected to give us a comprehensive insight into the deeper secrets of the Universe.
	In an overall classification we can say that GWs arise from two categories of relativistic astrophysical
	and cosmological sources \citep{Cai:2017cbj}. The former can be attributed to prevalent cases such as the compact binary inspirals, spherically asymmetric spinning neutron stars, pulsars (as well as other periodic sources), the explosion of supernovae, and other transient or burst sources. Note that the emission frequency band from these astrophysical objects is not expected to be higher than $10^4$ Hz. In relation to the latter category, there are a sequence of early Universe phase transitions (PTs) such as inflation, electroweak (EW), and quantum chromodynamics (QCD) which are potential candidates to create a notable stochastic background of the primordial GWs\footnote{Theoretically, cosmological sources that produce GWs is not restricted just to these three PTs rather there are other novel candidates such as the plasma instability \citep{Anand:2018mgf, Pandey:2019tmo, Pandey:2020gjy}, the photon-graviton conversion process \citep{Fujita:2020rdx} and/or the primordial black hole formation \citep{Nakama:2020kdc} in the early Universe which may have phenomenological  importance.} (see Refs. \citep{Caprini:2015zlo, Caprini:2018mtu, Mazumdar:2018dfl} for a detailed review particularly on the last two cases).
	Note that these PTs will result in the production of GWS if their nature is of strongly first order. More precisely, first-order PTs by creating true vacuum bubbles via the quantum tunneling process and subsequently the interaction with the environment of hot plasma, producing a stochastic background of GWs \citep{Kosowsky:1991ua,Kosowsky:1992vn,Kamionkowski:1993fg}. Due to the fact that the behavior of the stochastic GW spectrum is strongly sensitive to physics of the early Universe, it may carry footprints of the early cosmology which can be used to evaluate extended theories and quantum gravity models as well \citep{Bernal:2020ywq, Calcagni:2020tvw}.
	Since stochastic GWs (SGW) belong to very low frequency bands, highly sensitive detectors are needed to capture them. It is expected that in the coming years, one of the central aims of GW detectors will be the detection of SGWs. For instance, the frequency of SGWs arising from QCD-PTs is expected to be in the range of $10^{-9}-10^{-6}$ Hz. 
	A number of designed projects such as the European Pulsar Timing Array (EPTA) \citep{Kramer:2013kea}, the International Pulsar Timing Array (IPTA), a worldwide cooperation of astronomers involved in the analysis of PTA data \citep{IPTA):2013lea}, and the Square Kilometer Array (SKA) \citep{ska}, have the potential to detect the frequency range of the SGW.
	Actually, the idea of using PTAs, as natural interferometers with galactic-scale arm lengths,  in order to analyze GWs  was proposed in \citep{Hellings:1983fr} for the first time.
	Recently a new  analysis of 12.5 year PTA data by the North American Nanohertz Observatory for Gravitational Wave (NANOGrav) Collaboration \citep{Arzoumanian:2020vkk} found strong evidence of SGWs with a sensitivity of around 1 nHz - 10 nHz and $10^{-15}$ for frequency and GW strain amplitude, respectively. However, it seems that the sensitivity of NANOGrav in the measurement of strain amplitude is less than the projected sensitivity of SKA since it would be able to probe GW strain amplitudes around $10^{-17}$ \citep{Lazio:2013mea}. 
	Detailed information about some of these ongoing projects as well as other detectors e.g., evolved Laser Interferometer Space Antenna (eLISA) \citep{Seoane:2013qna}, Deci-hertz Interferometer GW Observatory (DECIGO) and Big Bang Observer (BBO) \citep{Yagi:2011wg} etc., with the capability of detecting higher frequency GWs in the range $10^{-6}$ Hz - 10 Hz can be found in~\citep{Moore:2014lga}.
	In general, the GWs, associated with different cosmological periods mentioned above, can shed light on the underlying physics related to the evolution of the Universe as specific features of the GW contain information about the concerned energy scale. Moreover, due to the very weak interaction GWs have the advantage of being protected from most of the environmental effects. This feature has made GWs very popular with cosmologists interested in exploring the early Universe, specifically before recombination (the interval between the Big Bang and the emission of CMB) when the environment for the propagation of electromagnetic waves was not transparent. However it is important to point out that according to the predictions of some theoretical frameworks it is also possible to have some cosmological and astrophysical sources that can generate
	GWs with frequencies higher than the range related to the above categories i.e., $10^{-9}$ Hz - $10^4$ Hz, see the recent white paper~\citep{Aggarwal} for details. This could be an incentive to research and develop detectors with a reach of even higher frequency than LIGO.
	In this paper we are particularly interested in the generation of GWs from the last PT experienced by the Universe which according to the standard model happened around $10^{-5}$s after the big bang in the temperature  range $T_*\sim 0.1-0.2$ GeV i.e., QCD-PT.
	In order to look for the stochastic background of GWs via QCD-PT, we  need to have accurate knowledge of the nature of the QCD-PT.
	However, determining the type of PT is one of the most serious challenges of both QCD-based cosmology and particle physics. Some nonperturbative lattice simulation studies reject the possibility of first-order, and even second-order, QCD-PT but insist on an analytic crossover \citep{Aoki:2006we, Bhattacharya:2014ara}, meaning that one should not expect GW production. This is not the whole story, rather it is demonstrated that if the neutrino chemical potential is sufficiently large\footnote{According to Ref.~\citep{Boyarsky:2009ix} in case of sterile neutrino playing the role of dark matter implies a large neutrino chemical potential.} but within the allowed bounds of big bang nucleosynthesis, the cosmic QCD transition is then expected to be of first-order \citep{Schwarz:2009ii}. There are some phenomenological models~\citep{Gorenstein:1981fa, Fogaca:2010mf}  that explicitly vote in favor of a strong first-order PT in QCD.
	It would be interesting to note that there are a few lattice simulations that under some conditions which commonly are not so realistic, imply a first-order PT in the QCD epoch. 
	For instance, Refs. \citep{Aoki:2004iq} and \citep{Lucini:2012wq} can be mentioned, in which the former refers to a bulk first-order PT with three quark flavor which is unphysical PT in some lattice formulations, while the latter refers to large $N_c$ pure gauge in the absence of contribution of quark flavors. It is worth mentioning that although known lattice simulations such as \citep{Aoki:2006we, Borsanyi:2010bp, Bazavov:2011nk, Bhattacharya:2014ara} insist on crossover nature of QCD-PT, they have done within some approximations, such as zero or negligible chemical potential, which if relaxed may change the nature of the transition. So we still do not know  the exact nature of PT in QCD epoch, and indeed  the issue is open and in progress.
	On a different note, the observed baryon asymmetry  arising from electroweak-baryogenesis mechanism can be explained via the first-order PT, rather than other types of PTs~\citep{Cohen:1993nk}.
	All in all, during these years, the idea of SGWs production from a  first-order QCD-PT and subsequently examining their chances of detection by relevant ongoing and future detectors (EPTA, IPTA, NANOGrav and SKA), has become very popular~\citep{Caprini:2010xv, Aoki:2017aws, Anand:2017kar, Brandenburg:2021tmp, Ahmadvand:2017tue, Ahmadvand:2017xrw, Chen:2017cyc, Cutting:2018tjt, Capozziello:2018qjs, Khodadi:2018scn, Li:2018oqf, Shakeri:2018qal, Hajkarim:2019csy, Davoudiasl:2019ugw, Dai:2019ksi, Ahmadvand:2020fqv}.
	
	One of the attractive aspects within the framework of strong interaction is the study of QCD-PT in the presence of strong external magnetic fields (MFs). The motivation of such studies usually come from several phenomenological contexts.
	
	The first motivation comes from the studies of ultrarelativistic heavy-ion collisions at the RHIC and LHC at CERN. In the last four decades, we have witnessed impressive activities to rebuild conditions akin to those shortly after the big bang, known as quark-gluon plasma (QGP). It is expected that two highly-charged ions, impacting with a little offset, are able to generate very large MFs of up to values $(1-30)m_\pi^2$ equal to $\simeq10^{14}-10^{16}$ Tesla at RHIC and LHC, respectively~\citep{Skokov:2009qp}. 
	Despite that the quantitative estimations performed from the classical viewpoint tell us that the lifetime of these MFs are just a small fraction of the lifetime of QGP~\citep{Kolb:2002ve}, the MF during its lifetime may be close to its maximum strength~\citep{Fukushima:2012xw}. Besides, it is shown that in the central region of the overlapping nuclei, the change of the spatial distribution of the MF in the transverse plane is very smooth and negligible~\citep{Voronyuk:2011jd}. With these backgrounds in mind about MFs, people commonly study the diverse aspects of QCD physics in a strong homogeneous and static MF, see~\citep{Gusynin:1994re, Lee:1997zj, Fukushima:2008xe, Kharzeev:2010gr, Andersen:2012zc, Fayazbakhsh:2012vr, Kharzeev:2013ffa, Fayazbakhsh:2013cha, Mamo:2013efa, Tuchin:2013ie, Haber:2014ula, Andersen:2014xxa, Sadooghi:2016jyf, Shao:2019hen} for instance. In recent years, due to the possibility of achieving strong MFs at RHIC and LHC, much attention has been paid to studying the role of the MF on the QCD thermodynamics. By employing lattice-simulation methods in the computation of the thermodynamic observables (such as pressure, energy and entropy density) the effects of background MF on the QCD equation of state (EoS) have been studied~\citep{Bali:2014kia}.
	Although the lattice simulations definitely have not ruled out the possibility of a first-order PT in the QCD epoch, the analysis performed in \citep{Bali:2014kia} still indicate that the QCD-PT is a crossover in the presence of an external effect such as a strong and homogeneous MF. However, a promising hint in favor of first-order QCD-PT in the analysis of \citep{Bali:2014kia} is that by going to high values of MF the interaction measure seems to increase in the crossover which might indicate that at very strong MFs, the transition converts to a first-order one. In other words, MF dependency of interaction measure shows that as the MF becomes stronger, the transition temperature reduces. Additionally, according to the standard QCD phase diagram at finite temperature  the crossover region of the PT belongs to large temperatures \cite{Stephanov:2007fk, Plumberg:2018fxo}. So a drop in the transition temperature due to the presence of an external strong MF, potentially can be the bearer of this message that in the end the nature of the PT might turn into first-order. A further indication in favor of a first-order QCD-PT in the presence of an external MF, one can mention the lattice analysis performed in \cite{DElia:2010abb}. There, by utilizing the reweighted plaquette distribution at the critical couplings with different values of MFs, the numerical analysis has shown that by going from zero or small MF to high values, a single peak distribution turns to a double peak, typical of a first-order PT.
	Further analysis indicated that a QCD-PT in the presence of a MF occurs at a slower rate since the transition temperature gets dropped \citep{Agasian:2008tb,Bali:2011qj, Ayala:2015lta, Endrodi:2015oba}.

	Second, according to some scenarios, the presence of MFs in the QCD era may be due to the former well-known PTs such as inflation, and EWPT \citep{Vachaspati:1991nm}. The latter is especially important in the sense that at the EW scale the electromagnetic field is directly influenced by the dynamics of the Higgs field without need nontrivial extensions of the particle physics standard model. In general, from a theoretical point of view, there were various candidates for generating MF before the QCD era at early universe, see~\citep{Grasso:2000wj} for a review.
	
	Third, some compact astrophysical objects like neutron stars, known as magnetars, are expected to have an extremely powerful MF in surface, of the order of $10^9-10^{11}$ Tesla~\citep{Duncan:1992hi, Kaspi:2017fwg}. Commonly, the magnetars are modeled as hybrid stars and include a core of hot QCD matter with a first-order PT.
	
	So, the MF is able to change the EoS and this has attracted people toward finding the modification of the EoS in the presence of a strong background MF via different descriptive frameworks. In this direction, the thermal QCD-EoS with a background of a strong and homogeneous MF has been studied within the effective quasi-particle description of QGP~\citep{Kurian:2017yxj, Koothottil:2018akg} and also perturbatively up to one-loop order~\citep{Rath:2017fdv,Karmakar:2019tdp}.
	Subsequently, this gives a theoretical motivation to explore the phenomenological effect of the presence of strong background MFs on the stochastic GW signal generated during the QCD-PT. Given that the exact nature of the QCD-PT is still debatable, this underlying scenario may seem speculative. However, adopting the first-order PT, lets us make a prediction on the role of an external MF on the stochastic background of GW.
	With this aim in mind we are going to revisit the energy spectrum calculation of GW on the basis of \citep{Karmakar:2019tdp} where utilizing one-loop approximation up to a specific order of coupling constant, the thermodynamic observables of a strongly magnetized hot QCD matter have been computed. This led to extracting the modified QGP-EoS.

	The layout of the rest of this article is as follows.
	In Sec. \ref{QCDEoS} we first overview hot QCD-EoS, with two light flavors, modified in the presence of background strong MF. In Sec. \ref{QCDmf}, by having the QGP-EoS, we then derive the general expression for the stochastic GW. In Sec. \ref{QCDS}, by discussing the three known processes during the first order PT that lead to GW generation, we are able to predict the response of the MF to stochastic GW detection in ongoing and proposed detectors for future observations. In Sec. \ref{con}, we give a summary of the results and conclude.
	
	%*************************************************
	%%%%%%%%%%%%%%%%%%%%%%%%%%%%%%%%%%%%%%%%%%%%%%%%%%%%%%%%%%%%%%%%
	\section{EoS for Ideal QGP in background strong magnetic field} 
	\label{QCDEoS}
\begin{figure*}
		\includegraphics[scale=0.4]{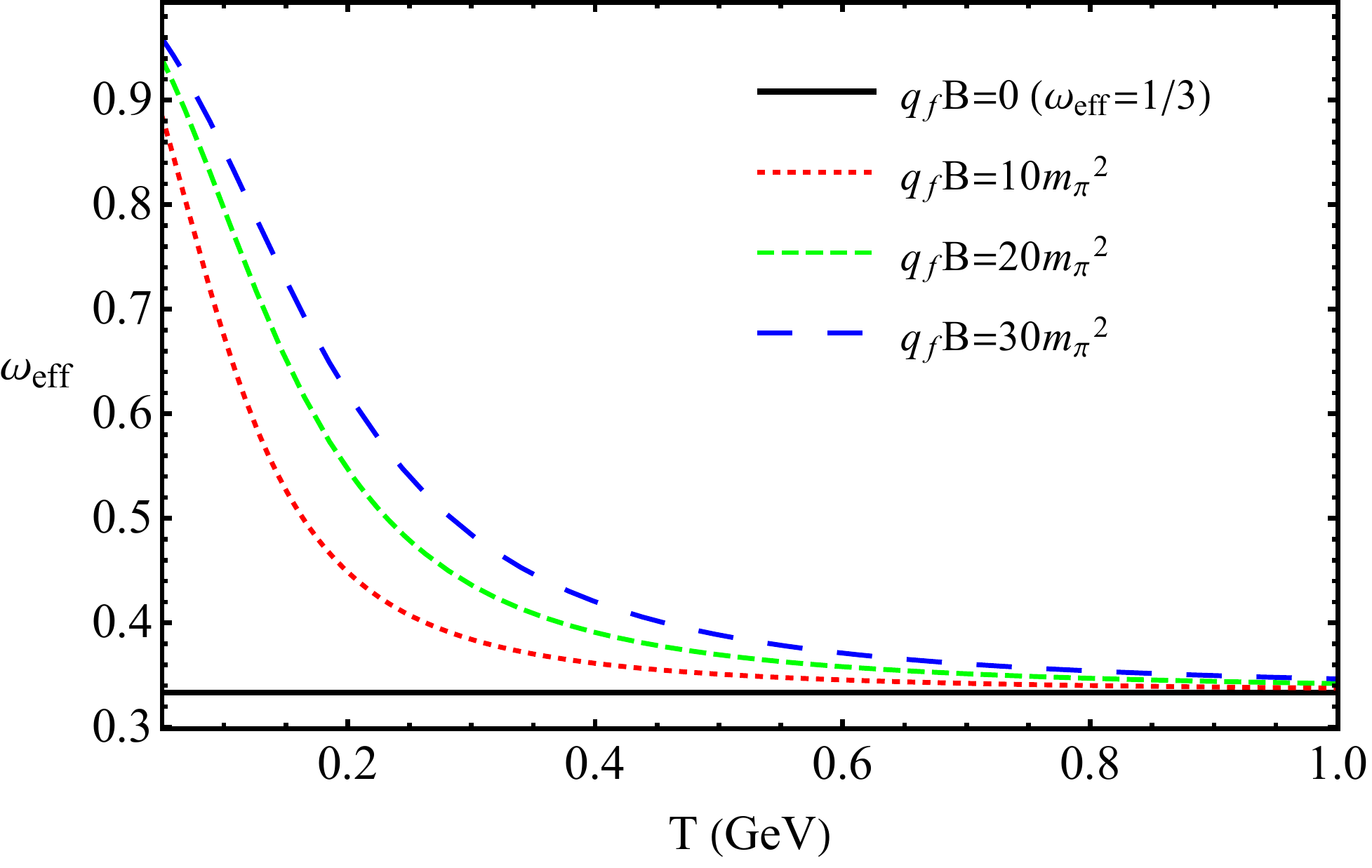}~~~~
		\includegraphics[scale=0.4]{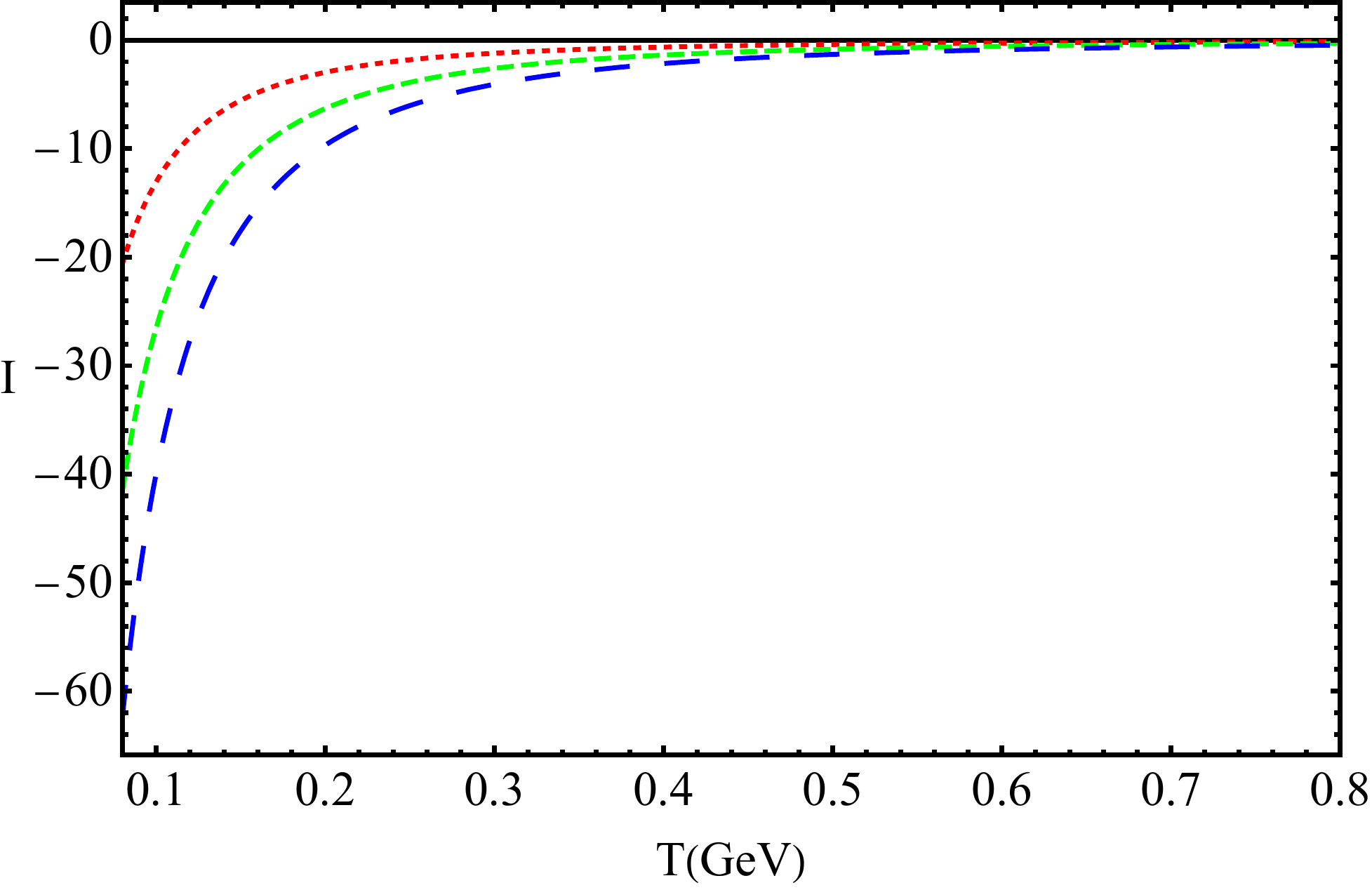}
		\caption{Variation of EoS parameter $\omega_{\text{eff}}$ and trace anomaly, in terms of temperature, respectively in the left and right panels. Here we used numerical values: $N_f=2,~N_c=3,~\gamma_E\simeq0.57721$  and $m_\pi^2\simeq0.02$ GeV$^2$.}
		\label{figWTR}
	\end{figure*}

	%%%%%%%%%%%%%%%%%%%%%%%%%%%%%%%%%%%%%%%%%%%%%%%%%%%%%%%%%%%%%%%%
	In this section, without going into much detail and focusing solely on the results of Ref.~\citep{Karmakar:2019tdp} we will briefly review the EoS of thermal QCD including $N_f$ quark flavors with $N_c$ colors and coupling constant $g$ in the presence of a strong static magnetic field directed along the $z$-axis ($\vec{B}=B\hat{z}$).
	The effects of MF is computed up to one-loop order in which the quark and gluon contributions are affected by the MF. The standard procedure is to calculate the free energy arising from quarks and gluons that are obtained via the functional determinant of the their effective one-loop propagators and then evaluate the pressure and other thermodynamic observables from the free energy. The renormalized final expressions of the quarks and gluons contributions in the free energy (as the starting point of deriving the relevant thermodynamic quantities) take the following forms~\citep{Karmakar:2019tdp},
	\begin{align}
		F_q^r = &- N_cN_f \sum_f\frac{q_f B T^2}{12} - 4N_c N_f \sum_f \frac{\(q_f B\)^2}{(2\pi)^2}\frac{g^2 (N_c^2-1)}{8 \pi^2 N_c}\Bigg[\frac{1}{24576}\Bigg\{12288 \ln 2\big(3 \gamma_E + 2\ln \frac{\Lambda}{2\pi T} + \ln 16\pi\big) \nn \\
		& +\frac{256\zeta[3]}{\pi^4T^2}\bigg(-\frac{3  g^2  \ln 2 (N_c^2-1)}{2N_c}q_fB + 6\pi^2 q_fB \ln  \frac{\Lambda}{2\pi T} +3 \pi ^2 q_fB\times(2+3 \gamma_E +\ln 16\pi )+2 \pi ^4 T^2\bigg)\nn \\
		&-\frac{4g^2(N_c^2-1)}{\pi^6N_cT^4}(q_fB)^2 \zeta[3]^2(4 + 105 \ln{2}) +\frac{7245\zeta[3]^3g^2(N_c^2-1)}{2\pi^8N_cT^6}(q_fB)^3
		\Bigg\}\Bigg], 
		\label{fq}   
	\end{align}
	and
	\begin{align}
		F_g^r=& \frac{N_c^2-1}{(4\pi)^2}\Bigg[
		- \frac{16 \pi^4 T^4}{45}+\frac{2N_c g^2\pi^2T^4}{9} + \frac{1}{12}\(\frac{N_c g^2T^2}{3}\)^2 \bigg(8 - 3 \gamma_E - \pi^2 + 4\ln 2 - 3 \ln\frac{\Lambda}{4\pi T}\bigg)\nn\\
		& + \frac{N_f\pi ^2 T^2}{2} \(\frac{g^2}{4\pi^2}\)^2  \sum_{f} q_fB\bigg(\frac{2\zeta'(-1)}{\zeta(-1)}-1+2\ln\frac{\Lambda}{2\pi T}\bigg) +\bigg(N_f^2+\sum_{f_1,f_2}\frac{q_{f_1}B}{q_{f_2}B}
		\bigg)
		%%%%%%%%%5
		\frac{g^4T^4}{32} \bigg(\frac{2}{3}\ln\frac{\Lambda}{4\pi T} -\frac{60  \zeta '[4]}{\pi^4} \nn \\
		& -\frac{1}{18} (25-12 \gamma_E -12 \ln 4 \pi ) \bigg) -\frac{1}{2}\(\frac{g^2}{4\pi^2}\)^2
		%%%
		\sum_{f_1,f_2} q_{f_1}B q_{f_2}B \bigg( \ln\frac{\Lambda}{4\pi T}+\gamma_E +\ln 2\bigg)\nn \\
		& -\frac{N_cN_f g^4T^4}{36} \bigg(1 - 2\frac{\zeta'(-1)}{\zeta(-1)} -2 \ln\frac{\Lambda}{4\pi T}\bigg) -\sum_f \frac{N_cg^4T^2q_fB}{144\pi^2} \bigg(\pi ^2-4+12\ln \frac{\Lambda}{4\pi T} - 2\ln 2\bigg(6\gamma_E + 4 \nn\\
		& + 3\ln2 -6\ln \frac{\Lambda}{4\pi T}\bigg) +12 \gamma_E  \bigg)\Bigg]-\frac{(N_c^2-1)g^3}{12\pi}\bigg( \frac{N_c T^2}{3}+\frac{q_fB}{4\pi^2}\bigg)^{3/2}T~,
		\label{fg}
	\end{align}
	respectively. Here $\gamma_{E},~\zeta$, and $\Lambda$ are respectively the Euler-Mascheroni constant, the Riemann zeta function, and the renormalization scale which usually is set as $\Lambda=2\pi T$. 
	At a first glance, it can be seen that the MF background gives a  contribution to the free energy of quarks which in the case of the absence of a MF, it disappears. 
	We note that the MF directed along a special axis breaks the translational invariance of space. As a consequence, the quark propagator becomes a function of separate elements of momentum, transverse ($P_{\perp}$) and longitudinal ($P_{\parallel}$) to the $B\hat{z}$.
	Now by discarding terms including $\mathcal{O}(g^2)$ or higher (since in the strong MF limit $q_fB\gg T^2$, the running coupling  constant is very small $g\ll1$ \footnote{Note that the QCD coupling $g$ will now run with the MF in addition to the temperature. Due to our interest in the strong MF limit i.e., $qB \gg T^2$, it is expected that by admitting the lowest Landau level (LLL) approximation to having the main contribution in physical quantities, the QCD coupling runs exclusively with the MF and is almost independent of the temperature. This is because the most dominant scale available is the MF and not the temperature.
		In the absence of a MF, the coupling constant $g$ is connected to the strong coupling constant $\alpha_s$ as $g=\frac{4\pi \alpha_s}{\Lambda}$ in which, due to competition between the contribution of the quarks and gluons in logarithmic relation of  $\alpha_s(k^2)$, thereby dependent on energy scales $k^2$, it decreases or increases \citep{Ferrer:2014qka}. However, in the presence of a MF, the strong coupling $\alpha_s$ splits into $\alpha_s^{\|}$ and $\alpha_s^{\perp}$, respectively parallel and perpendicular to direction of MF which only the former gives contribution to $g$. Besides, since $|q_fB|$ effectively acts as a cutoff in the LLL approximation, the coupling constant takes the form $g=\frac{4\pi \alpha_s^{\|}}{|q_fB|}$. As a result, in the strong field limit for consistency in LLL approximation, one ignores any temperature-dependency in $\alpha_s^{\|}$ such that it takes the form a logarithmic function of the MF as $\alpha_s^{\|}=\bigg(\frac{11N_c-2N_f}{12\pi}\ln \frac{|q_fB|}{\Lambda_{QCD}^2}\bigg)^{-1}$ \citep{Miransky:2002rp}. Given that $\alpha_s^{\|}$ is a monotonically decreasing function of the MF,  it is clear that in the strong MF regime, the coupling constant $g$ severely drops such that one can safely ignore the contribution of its higher powers i.e., $\mathcal{O}(g^2)\ll1$.}) the quark and gluon free energies (\ref{fq}) and (\ref{fg}) result in following expressions for the longitudinal pressures,
	\begin{eqnarray}\label{PP}
		\begin{aligned}
			{P_{\parallel}^q(T,q_fB)} &= \sum_f\frac{N_cN_f  q_f B T^2}{12} ~,  \\
			{P_{\parallel}^g}(T,q_fB) &= \frac{ \pi^2 (N_c^2-1)}{45}T^4~,
		\end{aligned}
	\end{eqnarray} 
	where $P_q$ and $P_g$ are the contributions from quarks and gluons, respectively.
	The transverse pressure is given by
	\begin{align}
		P_{\perp}(T,q_fB) = \left(P_{\parallel}^q(T,q_fB)
		+ P_{\parallel}^g(T,q_fB)\right)
		+ q_fB \frac{\partial \left(F_q^r(T,q_fB)
			+ F_g^r(T,q_fB)\right)}{\partial (q_fB)} ~,
	\end{align}
	where its final form reads as 
	\begin{eqnarray}\label{PT}
		%\begin{aligned}
		P_{\perp}(T,q_fB)=\frac{\pi ^2
			\left(N_c^2-1\right)}{45}T^4~.
		%\end{aligned}
	\end{eqnarray} 
	It is clear that the transverse pressure of a magnetized ideal QGP gas is independent of the magnetic field, just as we expect from ideal gluon pressure. Note that here the total pressure is $P=\sqrt{(P_{\parallel}^q+P_{\parallel}^g)^2+P_{\perp}^2}$.
	Using the thermodynamic relations $S(T,q_fB)=\frac{\partial P(T,q_fB)}{\partial T}$ and $\varepsilon(T,q_fB)=TS(T,q_fB)-P(T,q_fB)$, one arrives at the following modified expressions in the presence of a MF for entropy density and energy density,
	\begin{align}
		\label{S}
		S(T,q_fB) =\Bigg[&\left(
		\frac{2 \pi ^2 \left(N_c^2-1\right)}{90} T^4
		+\frac{N_c N_f q_fB}{18}T^2\right) \bigg(\frac{4\pi ^2}{90}
		\left(N_c^2-1\right) T^3
		+\frac{N_c N_f q_fB}{18}T
		\bigg) \nonumber \\
		& + \frac{4\pi^4 \left(
			N_c^2-1\right)^2}{2025}T^7
		\Bigg] \left(\left(
		\frac{\pi^2 \left(N_c^2-1\right)}{45}  T^4
		+\frac{N_c N_f q_fB}{18} T^2\right)^2
		+\frac{\pi^4 \left(N_c^2-1\right)^2 }{2025}T^8
		\right)^{-1/2},
	\end{align}
	and
	\begin{eqnarray}
		\label{E}
		%\begin{aligned}
		\varepsilon(T,q_fB) = \frac{25 N_c^2 N_f^2 (q_fB)^2 T^4 
			+ 40\pi^2 N_c \left(N_c^2-1\right) 
			N_f q_fB T^6+24 \pi^4 \left(N_c^2-1\right)^2 T^8}{90 \sqrt{25 N_c^2 N_f^2 (q_fB)^2 T^4 + 20\pi^2 N_c \left(N_c^2-1\right) 
				N_f q_fB T^6+8 \pi^4 
				\left(N_c^2-1\right)^2 T^8}}~,
		%\end{aligned}
	\end{eqnarray} 
	respectively. Note that Eqs. (\ref{PP}), (\ref{PT}), (\ref{S}), and (\ref{E}) actually address the thermodynamic quantities of an ideal QGP gas in the presence of an external MF.
	In order to probe the role of strong background a MF on the thermodynamics of an ideal QGP gas, in Fig. (\ref{figWTR}) we display the behavior of two importance quantities namely, the EoS parameter $\omega_{\text{eff}} = \frac{P(T,q_fB)}{\varepsilon(T,q_fB)}$ as well as the trace anomaly $I=\dfrac{P(T,q_fB)-3\varepsilon(T,q_fB)}{T^4}$ for different value of $q_fB$ with respect to the temperature. It is quite clear that by moving away from high temperatures to the critical one ($0.1-0.2$ GeV), the external MF results in a deviation of the behavior of QGP from the ideal gas (IG) i.e., $\omega_{\rm eff}=1/3$ and $I=0$.
	In what follows, we investigate the phenomenological effect of such a deviation from the IG (around the transition temperature $T_*$) due to the effect of the strong external MF on the generation of
	stochastic GWs arising from the QCD-PT.
	%
	
	%%%%%%%%%%%%%%%%%%%%%%%%%%%%%%%%%%%%%%%%%%%%%%%%%%%%%%%%%%%%%%%
	\section{GW spectrum revisited by in the presence of external MF}
	\label{QCDmf}
	%%%%%%%%%%%%%%%%%%%%%%%%%%%%%%%%%%%%%%%%%%%%%%%%%%%%%%%%%%%%%%%
	The imprint of a strong static MF during the QCD-PT can be searched for in the present observable SGW which is propagated from the moment of PT to the current time. Assuming that since the PT the Universe expanded adiabatically i.e., the entropy $s=Sa^3$  remained fixed ($\dot s/s=0\,$), we get the following relation from Eq. (\ref{S})
	\bea \label{Tdot}
	\frac{dT}{dt} = -H\bigg[3\left(\frac{d}{dT}S(T,q_fB)\right)^{-1}S(T,q_fB)\bigg]~,
	\eea
	which represents the time variation of temperature in the presence of a static magnetic field background.
	Using the Hubble parameter $H$ in Eq. (\ref{Tdot}) and taking into account the fact that the energy density of the GWs is inversely proportional to the fourth power of the scale factor ($\propto a^{-4}$) we obtain the following relations
	\be
	\label{scale}
	\frac{a_*}{a_0} = \exp \left[\int_{T_*}^{T_0} dT  \left\lbrace 3\left(\frac{d}{dT}S(T,q_fB)\right)^{-1}S(T,q_fB)\right\rbrace^{-1}\right]\,,
	\ee
	and
	\be
	\label{eq:rho-gw}
	\varepsilon_{\rm gw}(T_0) = \varepsilon_{\rm gw}(T_*) \, \exp \left[4\int_{T_*}^{T_0} dT  \left\lbrace 3\left(\frac{d}{dT}S(T,q_fB)\right)^{-1}S(T,q_fB)\right\rbrace^{-1}\right]\,,
	\ee 
	for the scale factor and energy density of the GWs, respectively. Note that from here onward the subscripts ``0'' and ``$\ast$'' respectively refer to the relevant quantities  today and at the PT epoch. Now, by defining the density parameter of GW in today's era $\Omega_{\rm gw} = \frac{\varepsilon_{\rm gw}(T_0)}{\varepsilon_{\rm cr}(T_0)}$ as well as using the relation $\frac{\varepsilon_{\rm cr}(T_*)}{\varepsilon_{\rm cr}(T_0)} = \left(\frac{H_*}{H_0}\right)^2$, we obtain the following relation,
	\be\label{eq:gw_den}
	\Omega_{\rm gw}=\Omega_{\rm gw*}\bigg(\frac{H_*}{H_0}\bigg)^2 \exp \left[4\int_{T_*}^{T_0} dT  \left\lbrace 3\left(\frac{d}{dT}S(T,q_fB)\right)^{-1}S(T,q_fB)\right\rbrace^{-1}\right]\,.
	\ee
	In order to derive the ratio between the Hubble parameter during PT and today, we use the continuity equation, $\dot\varepsilon = -3H\varepsilon \left(1+ \omega_{\rm eff}\right)$ where $\omega_{\rm eff} = P/\varepsilon$ is the effective EoS parameter. 
	One can now combine Eq. (\ref{Tdot}) with the continuity equation and integrate from some early time in the radiation dominated epoch with relevant temperature $T_r=10^4$ GeV to the PT epoch
	\be \label{energy}
	\varepsilon(T_*) = \varepsilon(T_r)\, \exp\bigg[\int_{T_r}^{T_{*}} dT \,3(1+ \omega_{\rm eff})
	\left\lbrace 3\left(\frac{d}{dT}S(T,q_fB)\right)^{-1}S(T,q_fB)\right\rbrace^{-1}\bigg]\,.
	\ee
	\begin{figure*}
		\includegraphics[scale=0.35]{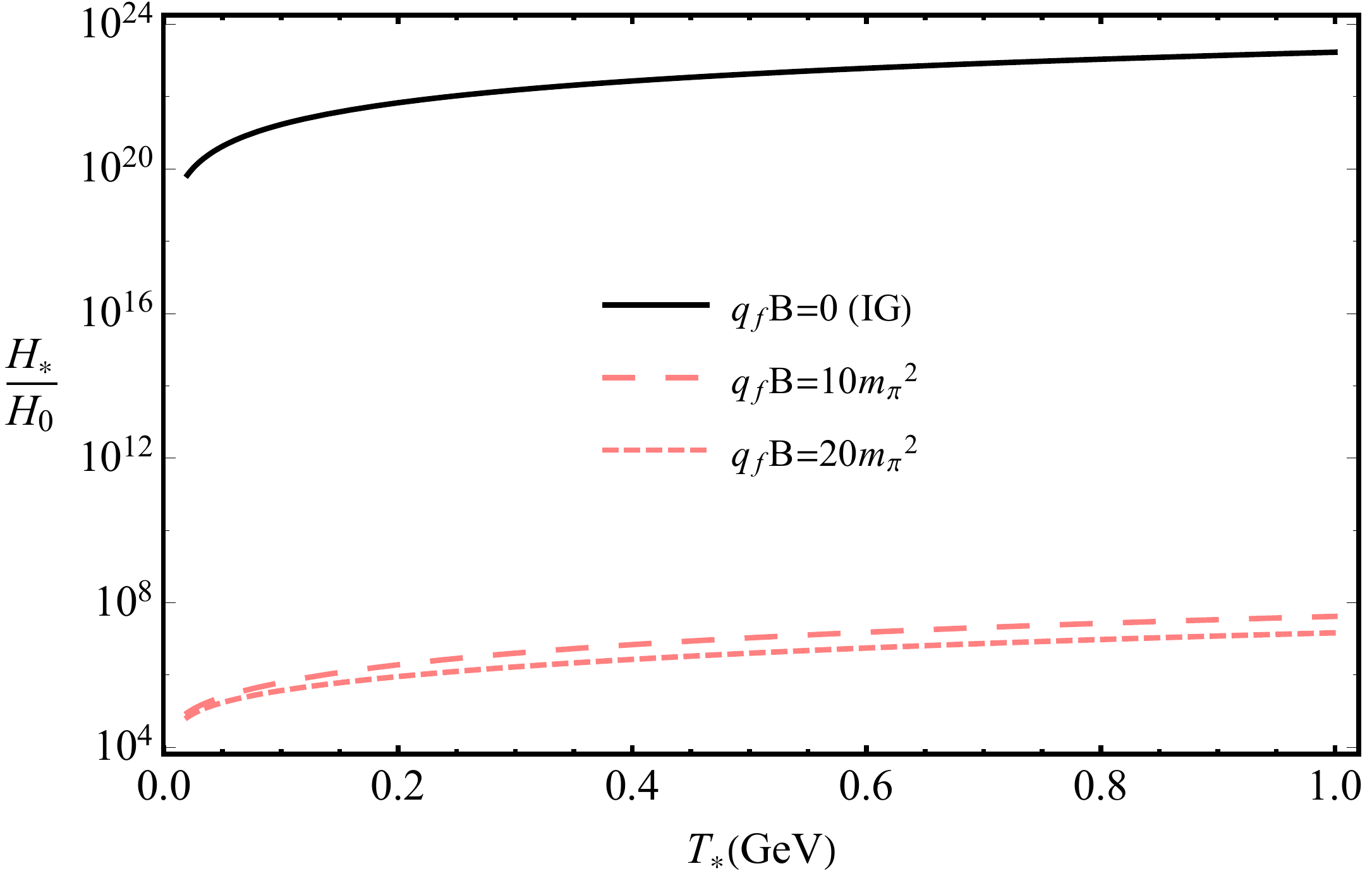}~~
		\includegraphics[scale=0.35]{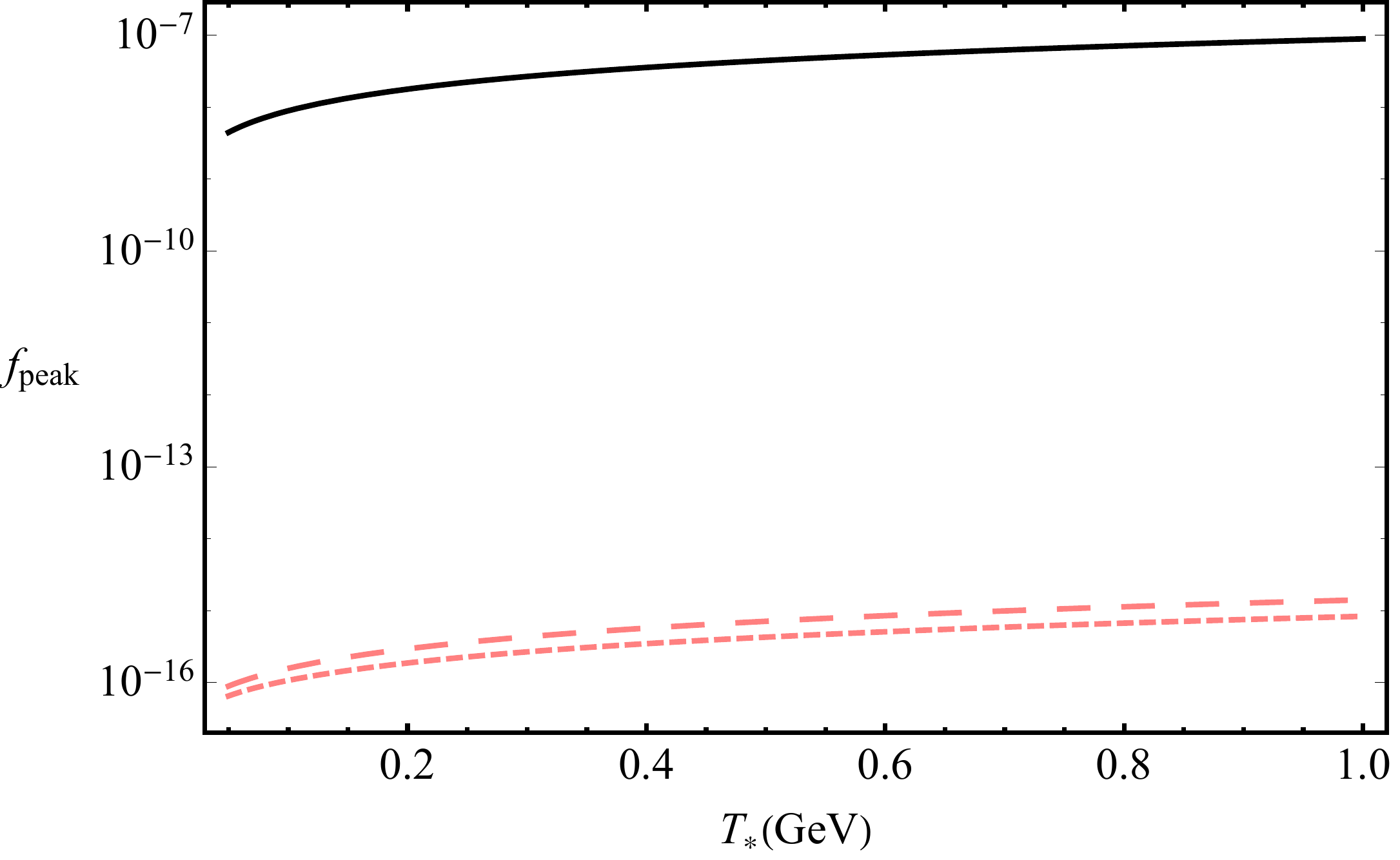}\\
		\includegraphics[scale=0.4]{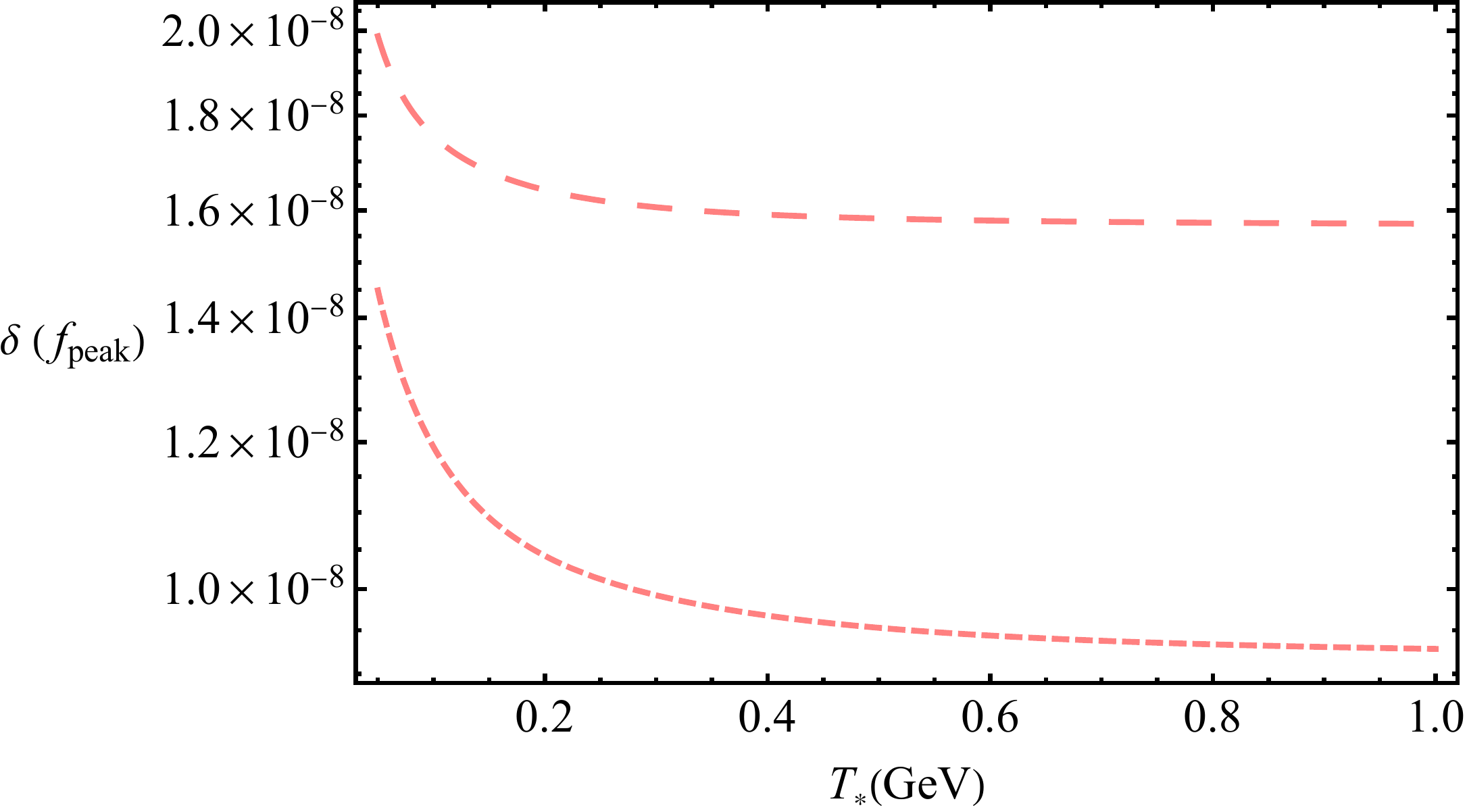}
		\caption{$H_*/H_0$, $f_{\rm peak}$ and also ratio $\delta(f_{\rm peak})=\frac{f_{\rm peak}(MF)}{f_{\rm peak}(IG)}$ as functions of PT temperature $T_*$. }
		\label{figHF}
	\end{figure*}
	As depicted in Fig. (\ref{figWTR}) in the presence of the MF, we no longer have the scenario where the ultrarelativistic particles obey  $\omega_{\rm eff}=\frac{P_q+P_g}{\varepsilon_q+\varepsilon_g}=\frac{1}{3}$. Despite that the value of $\omega_{\rm eff}$ asymptotically approaches the ideal value $1/3$, as temperature increases; however, with a drop in temperature it comes the critical point of the PT and the trace anomaly emerges so that QGP deviates from IG. In the following, we will see that this deviation from IG plays a considerable role on the peak frequency as well as the power of the GW signal produced during the QCD-PT. In the following we note that using the relation $H_*^2 \equiv \varepsilon_*$, Eq. (\ref{energy}) can be rephrased in terms of the Hubble parameter as,
	\begin{align}
		\label{eq:H*}
		\bigg(\frac{H_*}{H_0}\bigg)^2 = \Omega_{r0} & \exp \left[4\int_{T_0}^{T_r} dT  \left\lbrace 3\left(\frac{d}{dT}S(T,q_fB)\right)^{-1}S(T,q_fB)\right\rbrace^{-1}\right]
		\nonumber \\
		&\times \exp\bigg[\int_{T_{r}}^{T_{*}}
		dT~3(1+ \omega_{\rm eff} )
		\left\lbrace3\left(\frac{d}{dT}S(T,q_fB)\right)^{-1}S(T,q_fB)\right\rbrace^{-1}\bigg]\,,
	\end{align}
	where $\Omega_{r0}$ represents the today value of fractional energy density of radiation with a given value $\Omega_{r0}\simeq 8.5\times10^{-5}$. As a result, the GW spectrum measured today, Eq. (\ref{eq:gw_den}), takes the following form
	\begin{align}
		\label{eq:GW}
		\Omega_{\rm gw} = \Omega_{r0} \Omega_{\rm gw*}
		& \exp \left[4\int_{T_*}^{T_r} dT  
		\left\lbrace 3\left(\frac{d}{dT}S(T,q_fB)
		\right)^{-1}S(T,q_fB)\right\rbrace^{-1}\right]
		\nn \\
		& \times \exp\bigg[\int_{T_{r}}^{T_{*}}
		dT~3(1+ \omega_{\rm eff} )
		\left\lbrace 3\left(\frac{d}{dT}S(T,q_fB)\right)^{-1}S(T,q_fB)\right\rbrace^{-1}\bigg]\,.
	\end{align}
	Also, one can estimate the peak of the GW frequency\footnote{Peak frequency is actually the frequency of the signal at the time at which the strain amplitude (we will discuss this in the next section) reaches its peak. The importance of this quantity depends on the origin of the GW production. For instance, in two black holes mergers, it addresses information on the behavior of the system in the strong-field limit~\citep{Carullo:2018gah}. Here it may contain information about relevant issues to the QCD-PT.} redshifted to today via the following relation 
	\be 
	\label{eq:frequency}
	f_{\rm peak}=\left(\frac{a_*}{a_0}\right)= f_{*} \exp \left[\int_{T_*}^{T_0} dT  \left\lbrace 3\left(\frac{d}{dT}S(T,q_fB)\right)^{-1}S(T,q_fB)\right\rbrace^{-1}\right]\,.
	\ee
	Due to the direct relation between the frequency of gravitational radiation released at the period of PT ($f_*$) and the Hubble parameter at that time $H_*$, any decrease/increase in $H_*$ subsequently will alter the expected frequency of the gravitational radiation $f_*$. 
	By performing the integrals in Eqs. (\ref{eq:H*}) and (\ref{eq:frequency}) numerically one can find that taking the external MF into the EoS results in a considerable decrease of $H_*$ during PT as well as the peak frequency being redshifted to the present time. This can be seen explicitly from Fig. (\ref{figHF}). It also implicitly indicates a decrease in the PT temperature due to a strong MF, which is generally in agreement with outputs released in \citep{Agasian:2008tb,Ayala:2015lta,Endrodi:2015oba}.
	So it seems that the presence of a strong MF in EoS,  rules out finding any peak in the range of the characteristic frequency that can be probed in the current detectors with the sensitivity even around nHz.
	This can be seen more clearly in the next section of our analysis.
	\begin{figure*}
		\includegraphics[scale=0.3]{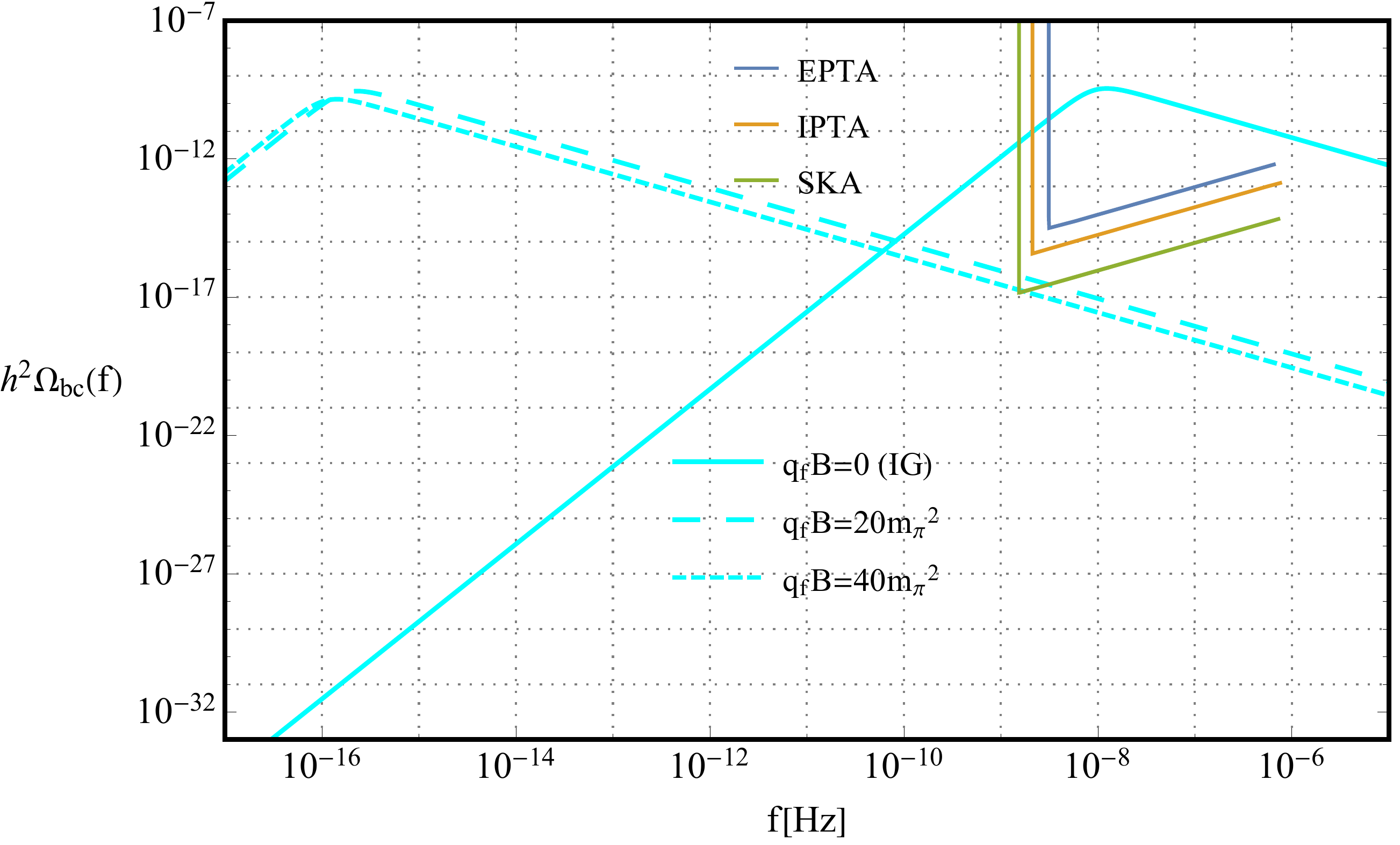}~
		\includegraphics[scale=0.3]{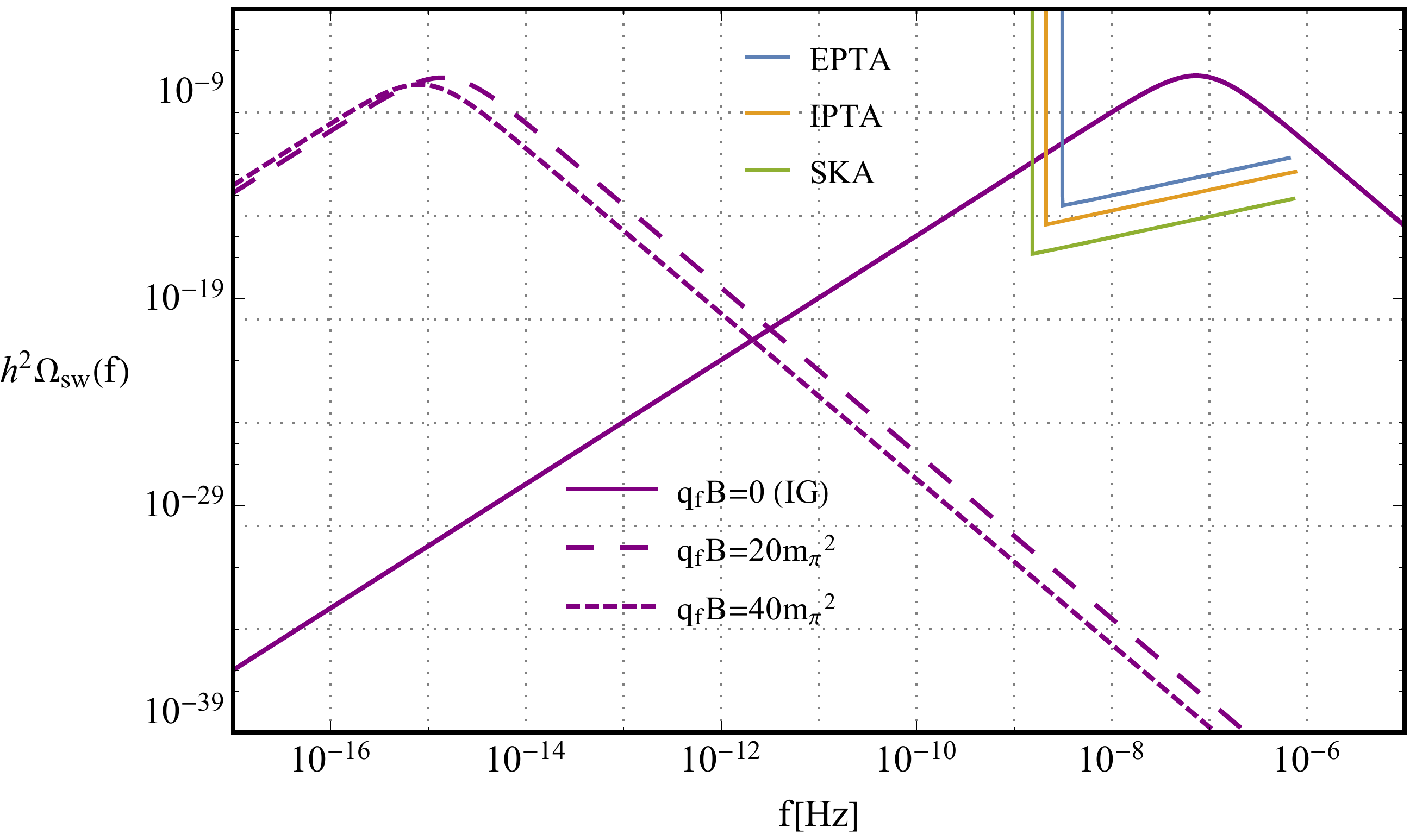}\\
		\includegraphics[scale=0.3]{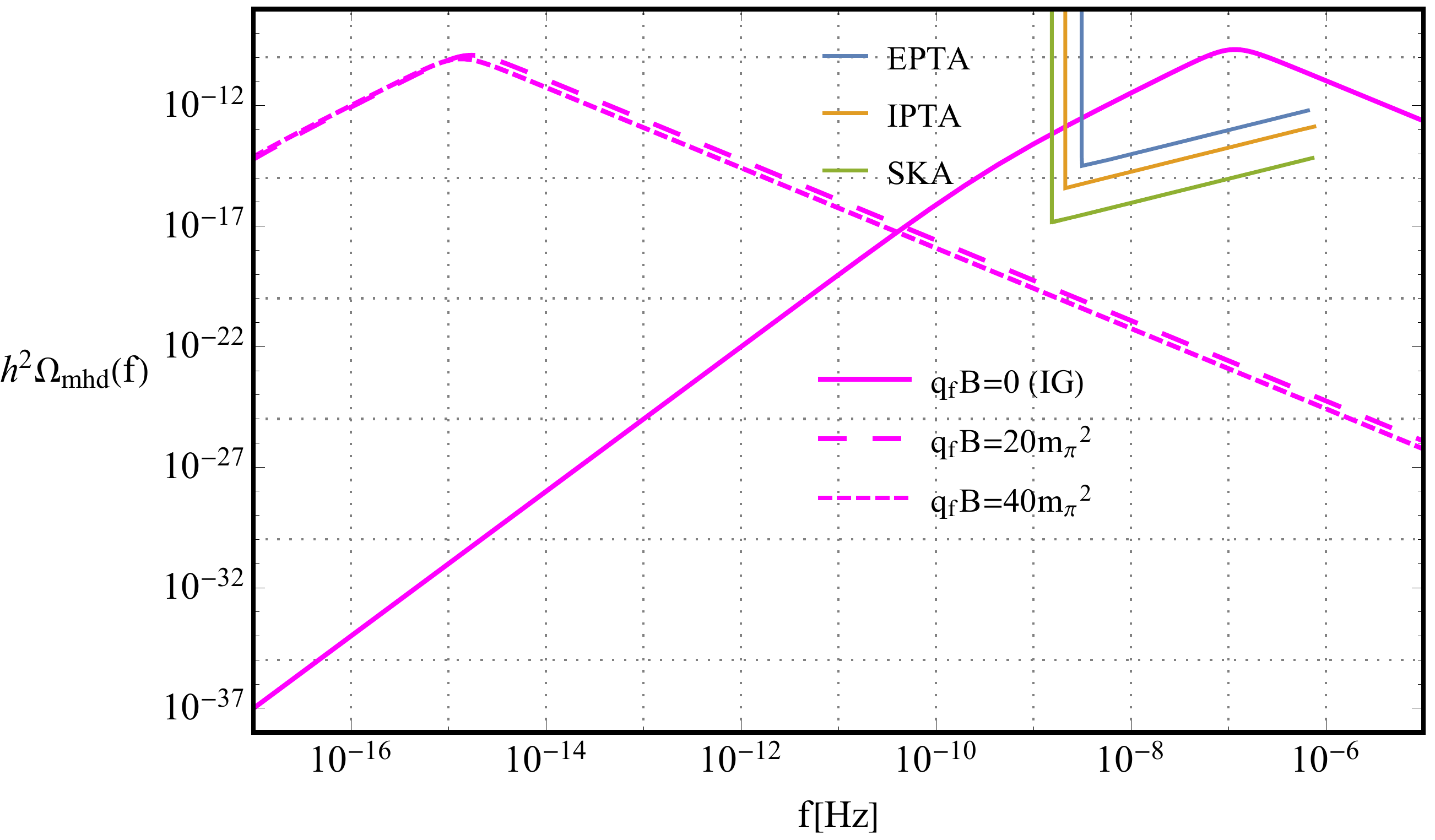}~
		\includegraphics[scale=0.3]{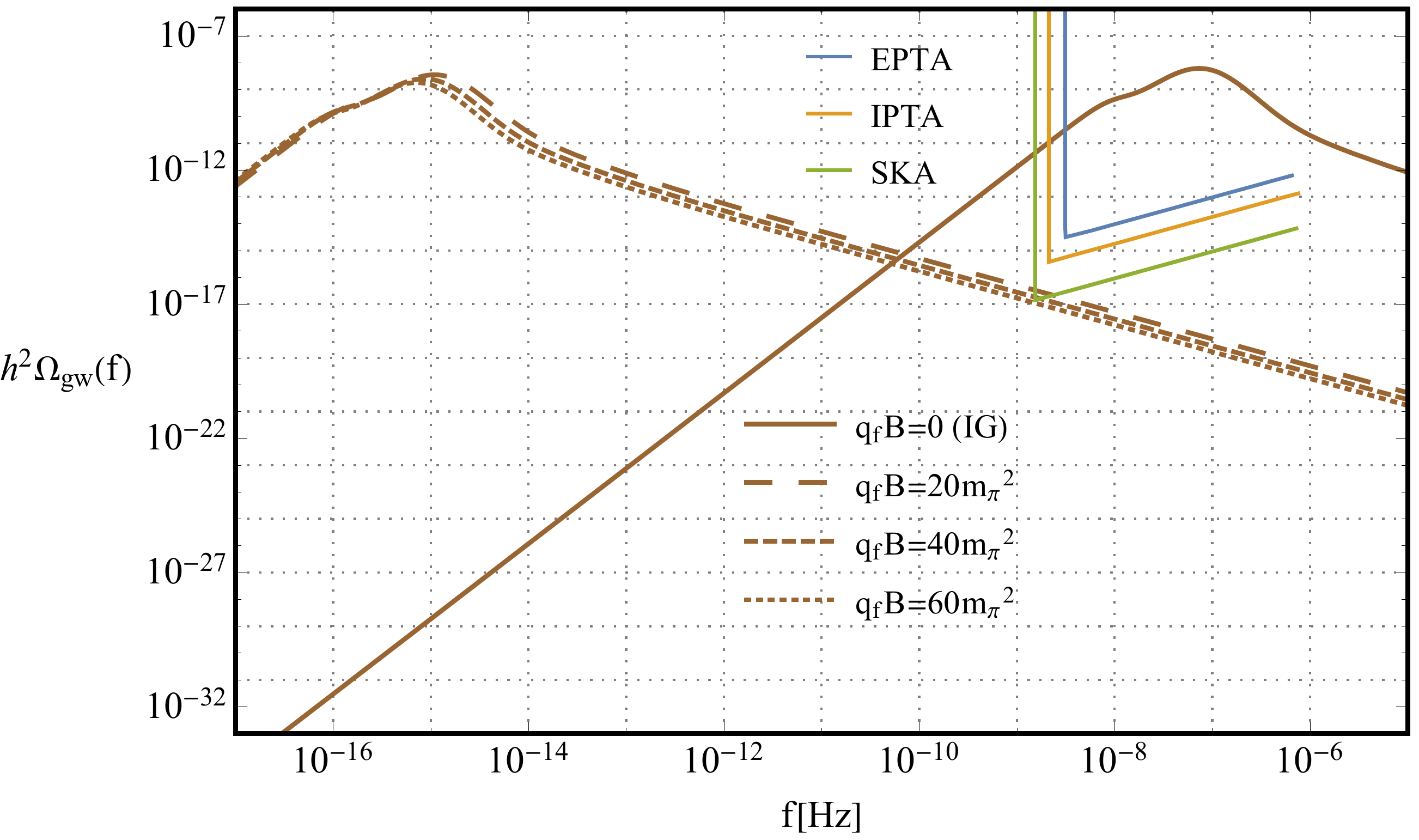}
		\caption{Comparing the stochastic background of the GW strain amplitudes arising from three sources:  BC, SW and MHD during the first-order QCD-PT in the absence/presence of an external strong MF with the projected reach of EPTA, IPTA and SKA. The net contribution due to all three sources to the GW signal, released in the right panel of the bottom row. Here, we set numerical values: $v_\mathrm{w}=0.7,~\beta=10H_*$, and $\chi_{\rm bc}=\chi_{\rm sw}=\chi_{\rm mhd}=0.05$.}
		\label{figS}
	\end{figure*}
	%EPTA \cite{Kramer:2013kea}
	%IPTA \cite{IPTA):2013lea}
	%SKA \cite{ska}
	%
	\begin{figure*}
		\includegraphics[scale=0.3]{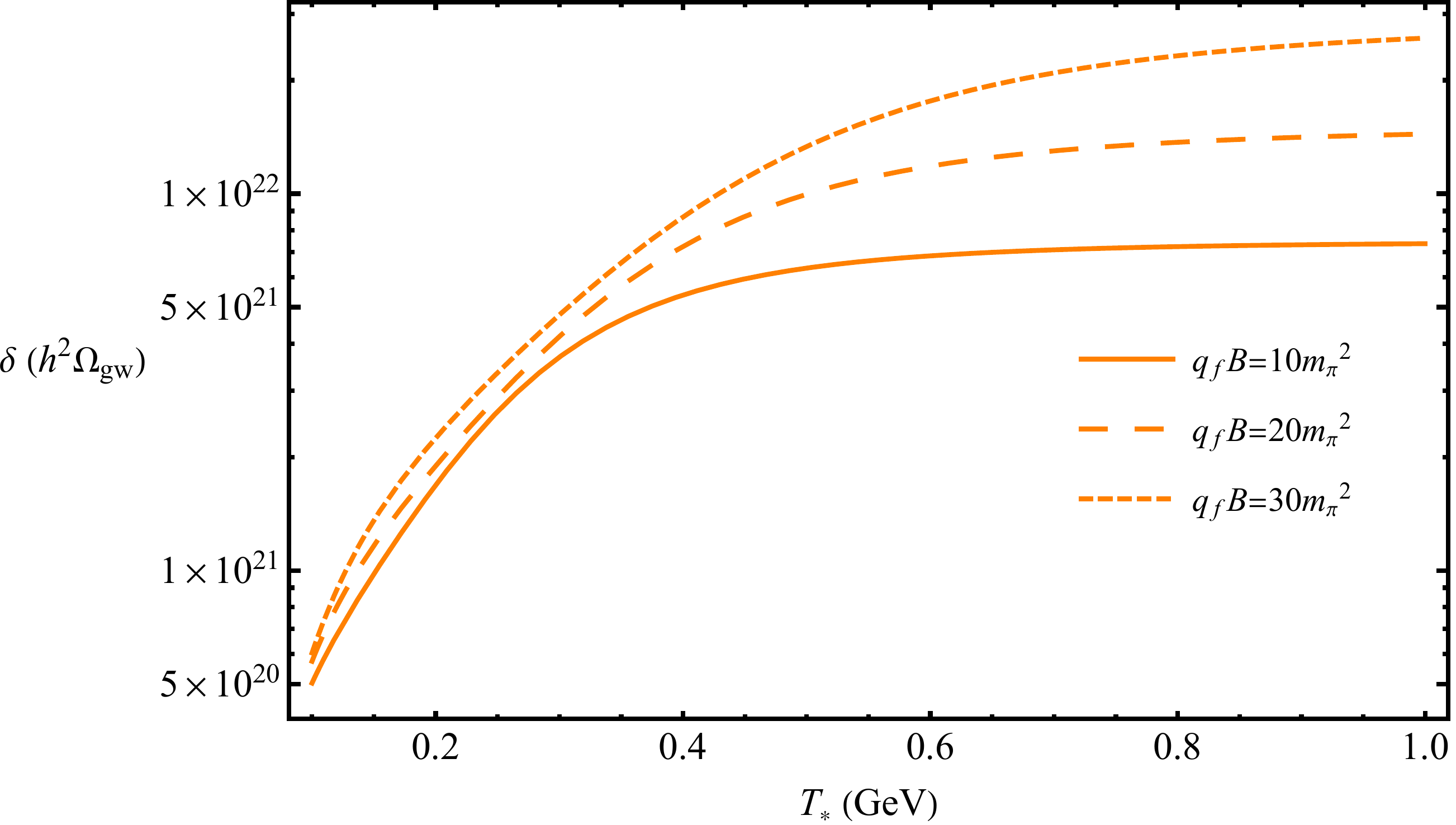}~~
		\includegraphics[scale=0.3]{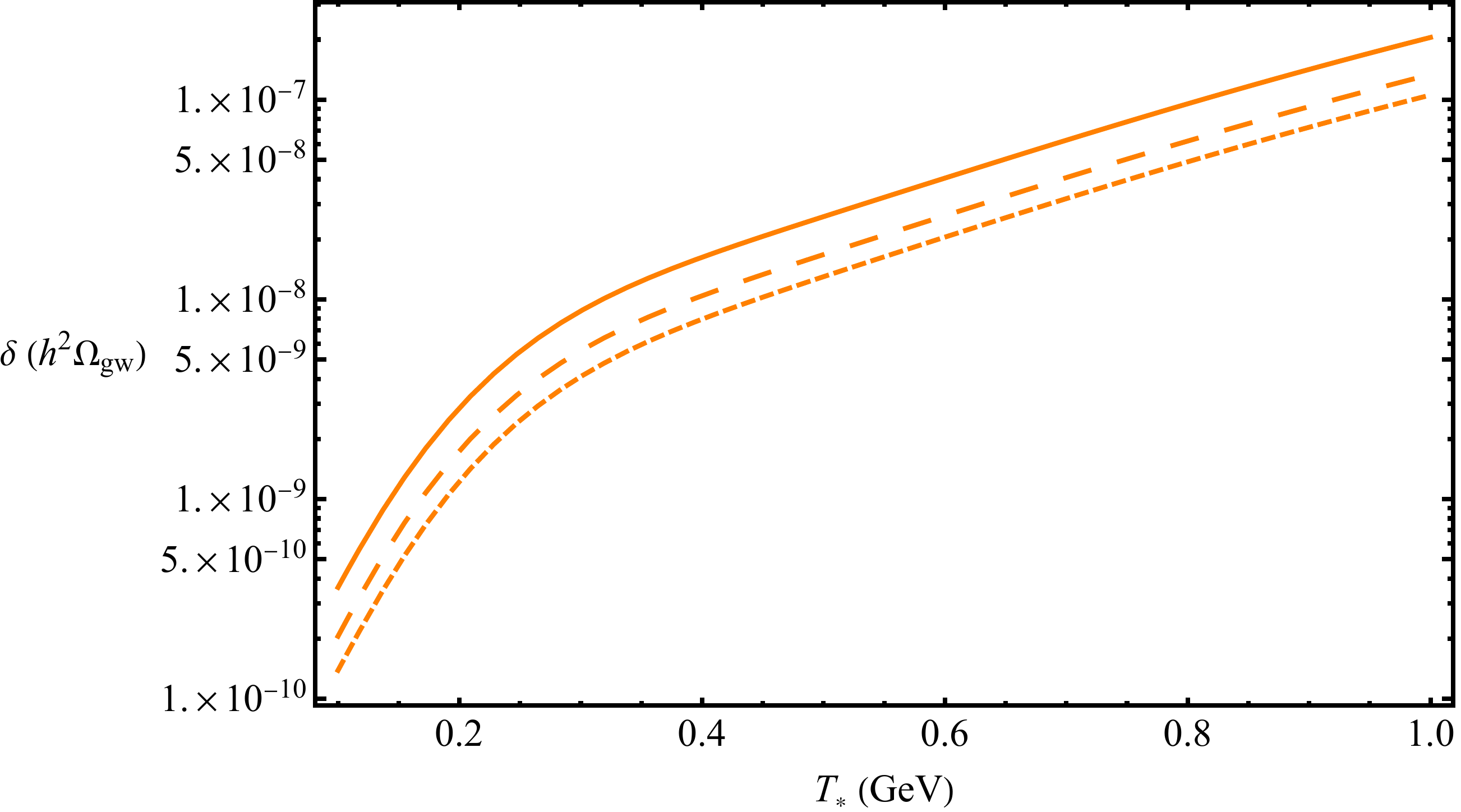}
		\caption{Ratio of the characteristic strain amplitude in the magnetized stochastic background of GW signal to its IG counterpart $\delta(h^2\Omega_{\rm gw})$ for two selected values of the peak frequencies $f_*=10^{-15}$ and $10^{-7}$ Hz in the left and right panels, respectively.}
		\label{figPer}
	\end{figure*}
	In passing we would like to mention that the ratio $H_*/H_0$ and the peak frequency $f_{\rm peak}$ change slightly by varying the value of $|q_fB|$ from $10M_\pi^2$ to $20M_\pi^2$ in Fig. (\ref{figHF}), while the change is large in comparison to the vanishing MF case $|q_fB|=0$. The reason for this apparent difference can be traced back to a nontrivial dependency of $|q_fB|$ in the integral  Eqs. (\ref{eq:H*}) and (\ref{eq:frequency}).
	
	%%%%%%%%%%%%%%%%%%%%%%%%%%%%%%%%%%%%%%%%%%%%%%%%%%
	\section{Contribution to the GW from QCD sources}
	\label{QCDS}
	%%%%%%%%%%%%%%%%%%%%%%%%%%%%%%%%%%%%%%%%%%%%%%%%%%
	In this section we discuss the role of strong magnetized EoS on the amplitude of the stochastic background of GWs generated during the QCD-PT in detail. During the first-order PT, the bubbles nucleate into the supercooling plasma and subsequently the latent heat related to the free energy between the symmetric and broken phases leads to the expansion of the bubble walls. Generally, in order to estimate the generated stochastic background of GWs via first-order QCD-PT, one should take into account the contributions of three well-known phenomena.
	\textbf{(a)} Bubble collisions (BC) in the context of first-order PT and GW production, first proposed by Witten \citep{Witten:1984rs}, comes subsequently after bubble nucleation and bubble expansion in order to start bubble percolation \citep{Huber:2008hg, Jinno:2016vai}. The mean radius of bubbles at collisions has an important role in the computation of the duration of PT. The GW power spectrum from the BC in the envelope approximation can be numerically fitted as $\Omega_\text{bc}(f)$ \citep{Huber:2008hg}
	\bea
	\label{env1}
	\Omega_\text{bc}(f) = \left(\frac{H_*}{\beta}\right)^2
	\chi_{\rm bc}^2 \left(\frac{0.11v_\mathrm{w}^3}{0.42+v_\mathrm{w}^2} \right) S_\text{bc}(f) \,,~~~~~~\chi_{\rm bc}=\frac{\kappa_{\rm bc} \alpha}{1+\alpha} 
	\eea
	where the spectral function $S_\text{bc}(f)$ has the following power-law form
	\bea
	\label{env2}
	S_\text{bc}(f) = \frac{3.8(f/f_b)^{2.8}}{1+2.8(f/f_b)^{3.8}},~~~~~~ f_b=\frac{0.62\beta}{1.8-0.1v_\mathrm{w}+v_\mathrm{w}^2}\,\frac{a_*}{a_0}
	\eea
	In the aforementioned relations, quantities $\alpha,~v_\mathrm{w}$, and $\kappa_{\rm bc}$ respectively denote the ratio between vacuum energy density produced during PT and radiation energy density, the expanding wall velocity; the coefficient which measures the fraction of the latent energy confined on the bubble wall during PT. Also $H_*/\beta$ is the nucleation rate relative to the Hubble rate during the PT which records the time duration of PT.
	\textbf{(b)} Sound waves (SW) are actually induced in the surrounding fluid due to the expansion of the bubbles so that in consequence of their collisions the SWs give rise to a nonzero tensor anisotropic stress that can be a rich source of GWs \citep{Hindmarsh:2015qta, Hindmarsh:2017gnf}. The power spectrum of GW due to SWs contribution, can take the power-law form as follows \citep{Hindmarsh:2015qta}
	\bea 
	\label{aw1}
	\Omega_\text{sw}(f) = \left(\frac{H_*}{\beta}\right)\chi_
	{sw}^2\,
	v_\mathrm{w} \, S_\text{sw}(f),~~~~~\chi_{\rm sw}=\frac{\kappa_\mathrm{sw}\alpha}{1+\alpha}
	\eea
	with the following function for the spectral shape,
	\bea 
	\label{aw2}
	S_\text{sw}(f) = \left(\frac{f}{f_\mathrm{sw}}\right)^3  \left(\frac{7}{4 + 3 (f/f_\mathrm{sw})^2 } \right)^{7/2}
	~,~~~f_\mathrm{sw}=\frac{38\beta}{31v_\mathrm{w}}\,\frac{a_*}{a_0}
	\eea
	The coefficient $\kappa_\mathrm{sw}$ in (\ref{aw1}) denotes the ratio of bulk kinetic energy to vacuum energy.
	\textbf{(c)} Apart from SWs, the bubble merging could also induce vortex motions in the surrounding fluid, which due to injection of energy arising from bubble collisions within the early ionized plasma with a very high Reynolds number, the magnetohydrodynamic (MHD) turbulence takes a form which is an independent source of GWs \citep{Caprini:2009yp, Binetruy:2012ze}. For the GW power spectrum due to the MHD turbulence component, we have \citep{Caprini:2009yp,Binetruy:2012ze}
	\bea 
	\label{tur1}
	\Omega_\text{mhd} (f) = \left(\frac{H_*}{\beta} \right) 
	\chi_{\rm mhd}^{\frac{3}{2}}\, v_\mathrm{w} S_\text{mhd}(f)~,~~~~~
	\chi_{\rm mhd}=\frac{\kappa_\text{mhd} \alpha}{1+ \alpha}
	\eea
	with the following function for the relevant spectral shape of the turbulent contribution,
	\bea
	\label{tur2}
	S_\text{mhd}(f) = \frac{(f/f_\text{mhd})^3}{[1+(f/f_\text{mhd})]^{\frac{11}{3}} ( 1+ 8 \pi f/h_*) }~,~~~ h_* =H_*\,\frac{a_*}{a}~,
	~~~f_\text{mhd} = \frac{7\beta}{4v_\mathrm{w}}\,\frac{a_*}{a_0}.
	\eea
	The coefficient $\kappa_\text{mhd}$ in (\ref{tur1}) measures the efficiency of the conversion of latent heat into turbulence. Note that there is no solid method to determine the value of $\kappa$ in the above functions. Hence, to facilitate the calculation we impose this assumption that $\chi_\text{bc}=\chi_\text{sw}=\chi_\text{mhd}$.
	As a result, the total GW power spectra can be expressed as the summation of above-mentioned three components
	\be 
	\label{eq:h}
	h^2\Omega_\text{gw}(f)=h^2\left(\Omega_{\rm bc}(f)+\Omega_{\rm sw}(f)+\Omega_{\rm mhd}(f)\right).
	\ee
	We are now in a position to investigate the detection possibility of the GW spectrum corresponding to BC, SW, and MHD sources in the presence of strong MF, through experiments such as EPTA, IPTA, and SKA that have arrayed to low-frequency detection of GWs. We show it in Fig. (\ref{figS}) for different values of free parameters, first separately for each source and then as their collective contribution.
	First, note that although we are able to draw these plots for different sets of free parameters $v_\mathrm{w},~\beta$, and  $\{\chi_{\rm bc}\,,\chi_{\rm sw}\,, \chi_{\rm mhd}\}$, however, in the overall behavior of curves we do not observe noticeable change. As is evident from Fig. (\ref{figS}) in the presence of strong MFs in the EoS we do not see any peak in the allowed range of characteristic frequency of detectors: EPTA, IPTA and SKA, while for QGP-IG there are peak frequencies around $10^{-8}$ Hz - $10^{-7}$ Hz. The peak frequencies observed here for the magnetic GW signals are justifiable since a MF in QGP-EoS causes the peak frequency to be redshifted to today, takes the frequencies much lower than IG-EoS, as shown already in Fig. (\ref{figHF}). A featured point in plots of Fig. (\ref{figS}) is that the magnetized EoS result in reduction of the GW strain amplitude so much that they appear below the permitted sensitivity of EPTA and IPTA. However, the expected signal within a certain range of the magnetic field ($\leq 40m_\pi^2$) can  cross off the detection range of SKA. It is also clear that among these three sources, the MF severely decreases the strain amplitudes related to sources SW and MHD as far as BC becoming dominant source in the net contribution to the GW signal. Another attractive point is that in the case of upgrading the frequency sensitivity of the current detectors to the fHz band, the magnetized GW signals can be probed, and instead the QGP-IG-based signal can be ruled out.
	In the end, for a clearer understanding of the effect of deviation from QGP-IG by an external strong MF on the strain amplitude of GW signal, in Fig. (\ref{figPer}) we show the ratio $\delta(h^2\Omega_{gw})=\frac{h^2\Omega_{gw}(MF)}{h^2\Omega_{gw}(IG)}$ in terms of critical temperature and different values of $q_fB$ and two selected values of the peak frequencies $10^{-15}$ and $10^{-7}$ Hz in the left and right panels, respectively.
	Two tips are evident from Fig. (\ref{figPer}) in overall agreement with Fig. (\ref{figS}). First, in the frequencies less than nHz, the strain amplitude of magnetized GW signal becomes stronger than its IG counterpart, while within the reach of these detectors at hand this is inverse. Second, as we expect by increasing the values of $q_fB$ in the left and right panels, the ratios of strain amplitude of the magnetized GW signal to its IG counterpart, become bigger and smaller, respectively.

	%%%%%%%%%%%%%%%%%%%%%%%%%%%%%%%%%%%%%
	\section{Conclusions}
	\label{con}
	%%%%%%%%%%%%%%%%%%%%%%%%%%%%%%%%%%%%%
	In this paper, by taking strongly magnetized QGP-EoS, we have investigated the contribution of magnetic field to the stochastic background of GWs arising from first-order QCD-PT.
	Actually, the presence of a magnetic field within EoS results in the appearance of a trace anomaly, meaning that we no longer deal with a QGP-ideal gas. This deviation from the QGP-ideal gas may leave some imprints on the stochastic background of the GW. With this consideration we have estimated two phenomenological quantities: peak frequency redshifted to today ($f_{\rm peak}$) as well as GW strain amplitude ($h^2 \Omega_{\rm gw}$).
	The former shifts dramatically to lower values than the ideal gas, so low such that one cannot expect it to be found  in the possible reach of characteristic frequency of the current detectors, even with the highest possible sensitivity around nHz band. This reduction could seem interesting from the perspective of the dynamic analysis of QCD-PT in the sense that it indicates a drop in PT temperature too.
	So, the strong magnetic field could mask the stochastic background of GWs produced by the QCD-PT. The latter, depending on the frequencies below and up to $\sim 1$ nHz, enhance and drop relative to the case of the ideal gas, respectively. By comparing the strain amplitudes corresponding to peak frequencies in the magnetized GW signals and the ideal gas counterpart, one finds that  they are of the same order of magnitude. This could be hopeful in the sense that merely by upgrading the frequency sensitivity of detectors in the future, the magnetized GW  is expected to be identified.
	We have found that in the presence of a strong magnetic field in hot QCD-EoS, the secondary sources SW and MHD are no longer able to play a significant role in generating GWs. Using the projected reach of detectors EPTA, IPTA, and SKA arrayed to detecting low-frequency GWs, we have shown that despite the absence of peak frequency for the magnetized signals in the currently available frequency range, in case of an upper bound for the magnetic field  ($q_fB\leq40m_\pi^2$) it is still possible for the signal tails to cross the SKA region.
	In summary, given the phenomenological importance of finding a peak around some frequencies in the GW spectrum  as identification of GWs with different origins, we should not expect the tail of magnetized stochastic background of GWs to leave any efficient signatures in the above-mentioned detectors. Indeed, the worthwhile information of the GW spectrum is encoded in the peak frequency of the signal rather than in its tail. So, this goal should be left to future generations of more sensitive detectors.
	
	\section*{Acknowledgements}
	We would like to thank Sampurn Anand and Fazlolla Hajkarim for useful discussions. U. K. D acknowledges the support from Department of Science and Technology (DST), Government of India under the Grant Reference No. SRG/2020/000283. G. L. thanks INFN and MIUR for support.

\end{document}